\documentclass[aip,apl,amsfonts,amssymb,amsmath,groupedaddress,twocolumn]{revtex4}

\usepackage{graphicx}
\usepackage{amsmath}
\usepackage{amsfonts}
\usepackage{amssymb}
\usepackage{graphicx}
\usepackage{epstopdf}

\begin{document}
\title{Steady-State Heat Transport: Ballistic-to-Diffusive with Fourier's Law}

\author{Jesse Maassen}
\email{jmaassen@purdue.edu}
\author{Mark Lundstrom}
\affiliation{Network for Computational Nanotechnology, School of Electrical and Computer Engineering, Purdue University, West Lafayette, IN 47907, USA}

\begin{abstract}
It is generally understood that Fourier's law does not describe ballistic phonon transport, which is important when the length of a material is similar to the phonon mean-free-path. Using an approach adapted from electron transport, we demonstrate that Fourier's law and the heat equation {\it do} capture ballistic effects, including temperature jumps at ideal contacts, and are thus applicable on all length scales. Local thermal equilibrium is not assumed, because allowing the phonon distribution to be out-of-equilibrium is important for ballistic and quasi-ballistic transport. The key to including the non-equilibrium nature of the phonon population is to apply the proper boundary conditions to the heat equation. Simple analytical solutions are derived, showing that {\it i)} the magnitude of the temperature jumps is simply related to the material properties and {\it ii)} the observation of reduced apparent thermal conductivity physically stems from a reduction in the temperature gradient and not from a reduction in actual thermal conductivity. We demonstrate how our approach, equivalent to Fourier's law, easily reproduces results of the Boltzmann transport equation, in all transport regimes, even when using a full phonon dispersion and mean-free-path distribution. 
\end{abstract}

%\pacs{85.65.+h, 73.63.-b, 72.20.Dp}
% 81.07.Bc Nanocrystalline materials
%73.50.-h Electronic transport phenomena in thin films
%31.15.Ac Ab initio group of atoms and clusters
%31.15.Ar Ab initio calculations
%81.07.Nb Molecular Nanostructures
%81.07.Lk Nanocontacts
%85.65.+h Molecular electronic devices
%72.10.Bg General formulation of transport theory
%72.20.Dp General theory, scattering mechanisms of conductivity
%73.23.-b Electronic transport in mesoscopic systems
%73.40.Sx Metal-semiconductor-metal structures
%73.63.-b Electronic transport in mesoscopic or nanoscale materials and structures

\maketitle

\section{Introduction}
\begin{figure}	
\includegraphics[width=6.5cm]{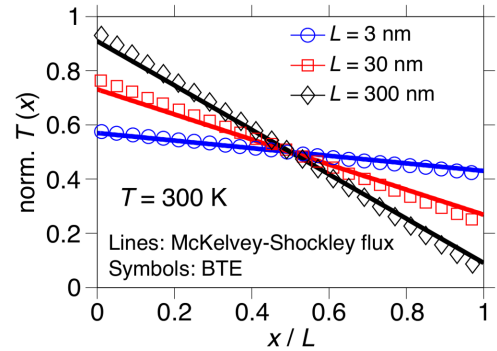}
\caption{Normalized temperature profile $(T(x)-T_R)/(T_L-T_R)$ versus normalized position $x/L$ for a Si film of $L=3,\,30,\,300\,{\rm nm}$. Symbols are results of the phonon BTE (taken from \cite{Escobar2008}). Lines are solutions to the approach described in this paper, which are obtained by solving a very simple phonon BTE or, equivalently, by solving $\nabla^2 T = 0$ with physically correct boundary conditions.} \label{fig1}
\end{figure}

Thermal transport at the nanoscale is a problem of great fundamental and practical interest \cite{Cahill2014}. Figure \ref{fig1} shows the temperature profiles across silicon films of varying length ($L$), as computed by the phonon Boltzmann Transport Equation (BTE) \cite{Escobar2008}. The temperature jumps at the interfaces with the two ideal, reflectionless contacts are characteristic features of quasi-ballistic phonon transport, and are commonly observed \cite{Majumdar1993,Chen2002} in physically-detailed modeling (such as Monte Carlo simulations\cite{Mazumder2001,Lacroix2005}, molecular dynamics \cite{Gomes2006,Hu2009,Sullivan2013}, or solutions of the phonon BTE \cite{Mahan1988,Claro1989,Majumdar1993,Joshi1993,Chen2002,Escobar2008,Hua2014,Vermeersch2014a,Vermeersch2014b}). In practical situations, the routine analysis of nanoscale heat transport phenomena, including ballistic effects, has been limited by the high computational demand of rigorous simulations. Simple, accurate and physically transparent models that provide physical insight could help in understanding the results of detailed simulations, as well as the analysis of experiments.

Quasi-ballistic phonon transport can impact the thermal response of materials on the nanoscale and even on the microscale, given that phonon mean-free-paths (MFP) can span from $\sim \!\!1\,{\rm nm}$ to $>\!\!10\,{\rm \mu m}$ \cite{Henry2008}. Ballistic phonon effects reduce the heat carrying capability of thin-films from the value expected from a simple application of Fourier's law \cite{Sellan2010,Bae2013}. This also affects the analysis of experiments probing short time and length scales (e.g., time/frequency-domain thermoreflectance) \cite{Koh2007,Regner2013}, influences heating in small electronic devices \cite{Cahill2014}, and can provide a route to extract the MFP distribution of materials \cite{Minnich2012,Yang2013}. A simple, physically sound, accurate, and computationally efficient technique to analyze such problems is presented in this paper. The lines in Fig. \ref{fig1} are the results of our calculations described below.

In this paper, we begin with an approach originally introduced by McKelvey \cite{Mckelvey1961} and Shockley \cite{Shockley1962} to describe particle transport, and extend it to treat heat transport. This technique uses a particularly simple, but accurate, discretization of the BTE into forward and reverse fluxes that provide solutions from the ballistic to diffusive limits. Additionally, we show that the McKelvey-Shockley phonon BTE can be recast exactly as Fourier's law and the heat equation with no additional assumptions (such as restrictions on the size of the structure or assumption of local thermal equilibrium). When solved with physically meaningful boundary conditions, the solutions are identical to those of the McKelvey-Shockley phonon BTE. The lines in Fig. \ref{fig1} are solutions of the heat equation, $\nabla^2 T\! =\! 0$, using the correct physical boundary conditions, as described in this paper.

The outline of this paper is as follows. Section \ref{sec:theory} describes our approach to heat transport. A simple example calculation is presented in Section \ref{sec:example}, where we show that the magnitude of the temperature jump is proportional to the phonon transmission across the structure. Section \ref{sec:fourier} provides a derivation of familiar heat transport equations from the McKelvey-Shockley phonon BTE. Although familiar in form, the heat equation includes ballistic effects \emph{if proper boundary conditions are used}. Section \ref{sec:discussion} discusses the treatment of a full phonon dispersion with energy-dependent mean-free-path, and explains how the calculations shown in Fig. \ref{fig1} are performed. Finally, we summarize our findings in Section \ref{sec:summary}.

\section{Theoretical Approach} 
\label{sec:theory}
\begin{figure}	
\includegraphics[width=8.5cm]{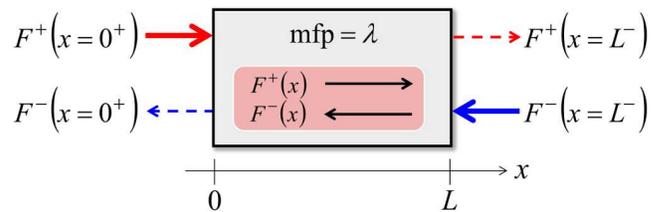}
\caption{Thermal conductor of length $L$ with a given mean-free-path for backscattering $\lambda$. By specifying the injected phonon fluxes (solid arrows), the McKelvey-Shockley flux equations describe the evolution of the flux components inside the material.} \label{fig2}
\end{figure}

For this work, we borrow an approach originally developed for electronic transport that is applicable on all length scales, the McKelvey-Shockley flux method \cite{Mckelvey1961,Shockley1962}, and adapt it for phonon/heat transport. We assume steady-state 1D transport along $x$ with an infinite $y$-$z$ plane. The essence of this approach is to {\it (i)} describe phonons in terms of fluxes (i.e. phonon density times average velocity along the transport direction) and {\it (ii)} categorize all phonons into two components: forward and backward fluxes. The governing equations of the McKelvey-Shockley flux method are \cite{Rhew2002}:
\begin{align}
\frac{{\rm d} F^+(x,\epsilon)}{{\rm d} x} &= -\frac{F^+(x,\epsilon)}{\lambda(\epsilon)}+\frac{F^-(x,\epsilon)}{\lambda(\epsilon)}, \label{mk_flux1} \\
\frac{{\rm d} F^-(x,\epsilon)}{{\rm d} x} &= -\frac{F^+(x,\epsilon)}{\lambda(\epsilon)}+\frac{F^-(x,\epsilon)}{\lambda(\epsilon)}, \label{mk_flux2}
\end{align}
where $F^+$/$F^-$ are the forward/backward phonon fluxes [units: \#phonons m$^{-2}$s$^{-1}$eV$^{-1}$], $\lambda(\epsilon)$ is the mean-free-path for backscattering and $\epsilon$ is the phonon energy. The above coupled equations describe the evolution of each flux type, which can scatter to/from the opposite flux component. $\lambda(\epsilon)$ governs the scattering, and is defined as the average distance travelled along $x$ by a phonon with energy $\epsilon$ before scattering into an opposite moving state (in the isotropic case, $\lambda$ is (4/3) times the regular MFP \cite{Jeong2010}). Note that the McKelvey-Shockley flux method can be derived from the BTE \cite{Rhew2002}. The boundary conditions are: 
\begin{align}
F^+(x=0^+,\epsilon) &= F^+_0(\epsilon) \label{flux_bound1}, \\
F^-(x=L^-,\epsilon) &= F^-_L(\epsilon) \label{flux_bound2},
\end{align}
where $L$ is the length of the thermal conductor. Thus one needs to specify the injected phonon fluxes at both ends: $F^+$ on the left side ($x=0^+$) and $F^-$ on the right side ($x=L^-$). The McKelvey-Shockley flux method described by Eqns.(\ref{mk_flux1}-\ref{mk_flux2}), with the boundary conditions given by Eqns.(\ref{flux_bound1}-\ref{flux_bound2}), forms the basis for our approach to heat transport (see Fig. \ref{fig2}).

The total heat current ($I_Q^{\rm tot}$) and heat density ($Q^{\rm tot}$) are written as:
\begin{align}
I_Q^{\rm tot} &= \int_0^{\infty}\epsilon\,\left[F^+(x,\epsilon)-F^-(x,\epsilon)\right]\,{\rm d}\epsilon,\;\;\;\text{[W\,m$^{-2}$]} \label{def:heat_curr} \\
Q^{\rm tot}(x) &= \int_0^{\infty}\epsilon\,\left[\frac{F^+(x,\epsilon)+F^-(x,\epsilon)}{v_x^+(\epsilon)}\right]\,{\rm d}\epsilon,\;\;\;\text{[J\,m$^{-3}$]} \label{def:heat_den}
\end{align}
where $F=F^+-F^-$ is the net phonon flux, $(F^++F^-)/v_x^+$ is the phonon density and $v_x^+$ is the average $x$-projected velocity (defined as $v_x^+=\sum_{k,v_x>0} v_x \delta(\epsilon-\epsilon_{k})/\sum_{k,v_x>0} \delta(\epsilon-\epsilon_{k})$). The heat current and heat density correspond to multiplying the net phonon flux and the phonon density, respectively, by the energy $\epsilon$ carried by each phonon (and integrating over all phonon energies). From the above definitions, we can directly replace $F^{\pm}$ in the McKelvey-Shockley flux equations (Eqns.(\ref{mk_flux1}-\ref{mk_flux2})) by $I_Q^{\pm}=\epsilon F^{\pm}$, as we will assume from this point on unless otherwise stated. In addition, to ease the notation we will drop the explicit dependence on $\epsilon$ in $I_Q^{\pm}$, $\lambda$ and $v_x^+$, although keep in mind that a final integration over energy is required (Eqns.(\ref{def:heat_curr}-\ref{def:heat_den})).

Equations (\ref{mk_flux1}) and (\ref{mk_flux2}) comprise a simple BTE in which the forward and reverse fluxes have been integrated over angle. This particular discretization is especially effective in handling the correct physical boundary conditions, where a flux is injected from each side. Inside the device, the carrier distribution can be very far from equilibrium, but each half of the distribution is at equilibrium with its originating contact as it is injected and scattering gradually mixes both flux components. In the limiting case of purely ballistic transport, each half of the distribution is in equilibrium with its originating contact. More complicated discretizations are possible and sometimes necessary, but we will demonstrate the effectiveness of these simple equations in the remainder of the paper.

\begin{figure}	
\includegraphics[width=7.5cm]{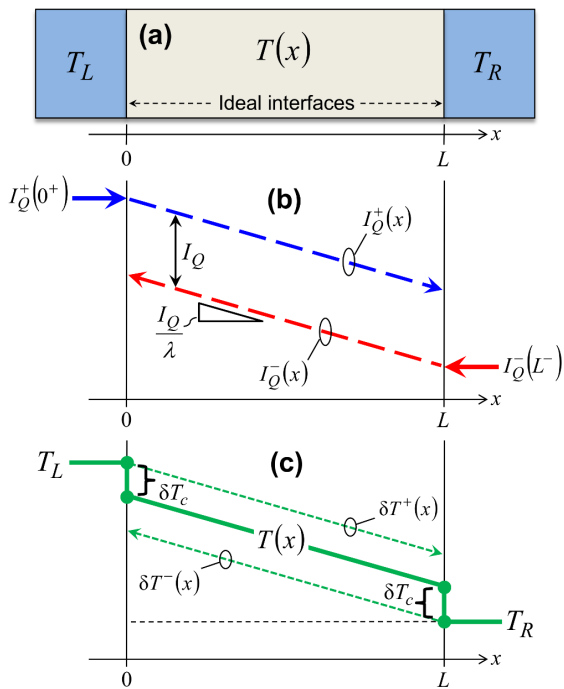}
\caption{(Color online) (a) System under study: a thermal conductor of length $L$ joined by two contacts in equilibium, each at a specified temperature $T_L$ and $T_R$. The contact interfaces are reflectionless and do not scatter phonons. (b) Forward and backward heat currents versus position $x$. (c) Temperature profile versus position $x$, which shows the temperature jumps ($\delta T_c$) at the contacts. The temperature profiles of the forward and backward phonons are shown as dashed lines.} \label{fig3}
\end{figure}

\section{Example: heat transport in a dielectric film}
\label{sec:example}
Having presented the McKelvey-Shockley flux method adapted for heat transport, we will now demonstrate this approach with an example. We will consider steady-state thermal transport across a dielectric film of length $L$ (the electronic contribution to thermal transport can be neglected), contacted by two {\it ideal} thermalizing contacts each at their respective temperatures $T_L$ (left contact) and $T_R$ (right contact), as shown in Fig. \ref{fig3}(a). ``Ideal contacts" in this context assumes that {\it (i)} each contact is in thermal equilibrium, with phonon statistics given by the Bose-Einstein distribution, and {\it (ii)} the interfaces are reflectionless, thus phonons are not scattered at the contacts. While perfect contacts are assumed in this paper, this is not a fundamental limitation of our approach, as we will discuss later.

We begin by subtracting the flux equations (\ref{mk_flux1}) and (\ref{mk_flux2}):
\begin{align}
{\rm d}I_Q/{\rm d}x &= 0. \label{cont_eq}
\end{align}
This is the energy balance equation, or equivalently the first law of thermodynamics, which states that under steady-state conditions the heat current $I_Q$ is a constant along $x$. Using this relation with Eqns.(\ref{mk_flux1}-\ref{mk_flux2}), we have 
\begin{align}
\frac{{\rm d}I_Q^+(x)}{{\rm d}x} &= \frac{{\rm d}I_Q^-(x)}{{\rm d}x} = -\frac{I_Q}{\lambda}. \label{def_fplusminus}
\end{align}
It is straightforward to show that $I_Q^+(x)$ and $I_Q^-(x)$ have the following solutions:
\begin{align}
I_Q^+(x) &= I_{Q,0}^+ - \frac{I_Q}{\lambda}\,x, \label{fplus} \\
I_Q^-(x)  &= I_{Q,L}^- - \frac{I_Q}{\lambda}\, (x-L), \label{fminus}
\end{align}
where we used the boundary conditions Eqns.(\ref{flux_bound1}-\ref{flux_bound2}). The forward and backward heat currents $I_Q^{\pm}$ vary linearly along $x$, as shown in Fig. \ref{fig3}(b). The difference between $I_Q^+$ and $I_Q^-$ is the net heat current $I_Q$, and the slope of $I_Q^+$ and $I_Q^-$ is $I_Q/\lambda$.

The net heat current $I_Q$ can be extracted by subtracting Eq. (\ref{fminus}) from Eq. (\ref{fplus}), and isolating $I_Q$:
\begin{align}
I_Q &= \left(\frac{\lambda}{\lambda+L}\right) \left[ I_{Q,0}^+ - I_{Q,L}^- \right], \label{fnet}
\end{align}
where $\lambda/(\lambda + L)$ is the phonon transmission coefficient $\mathcal{T}$, corresponding to the probability of a phonon traveling from one contact to the other. Thus, the net heat current is simply the difference in injected heat currents times the phonon transmission coefficient.

If the contacts are in equilibrium, then the injected heat currents from the contacts can be written as
\begin{align}
I_{Q,0}^+ &= \epsilon\,\frac{M}{h}\,f_{\rm BE}(T_L), \label{fleft} \\
I_{Q,L}^- &= \epsilon\,\frac{M}{h}\,f_{\rm BE}(T_R), \label{fright}
\end{align}
where $M(\epsilon)$ is the distribution of modes of the thermal conductor (depends only on the phonon dispersion) \cite{Maassen2013a,Maassen2013b}, $f_{\rm BE}$ is the Bose-Einstein occupation function and $h$ is Planck's constant. Inserting these expressions into Eq. (\ref{fnet}), we obtain:
\begin{align}
I_Q &=  \epsilon\,\frac{M}{h}\left(\frac{\lambda}{\lambda+L}\right) \left[ \, f_{\rm BE}(T_L) - f_{\rm BE}(T_R) \,\right] \label{fnet2} \\
      &\approx  \underbrace{\epsilon\,\frac{M}{h}\left(\frac{\lambda}{\lambda+L}\right) \frac{\partial f_{\rm BE}}{\partial T}}_{K} \, \Delta T, \label{fnet3}
\end{align}
where $K=K^{\rm ball}\,\mathcal{T}$ is the thermal conductance, $K^{\rm ball}$ is the ballistic thermal conductance \cite{Jeong2011}. Note that Eq. (\ref{fnet3}) applies in the case of a small $\Delta T$, while Eq. (\ref{fnet2}) does not have this limitation. The total heat current is calculated by integrating over energy (Eq. (\ref{def:heat_curr})). The above expressions for $I_Q$ are applicable on {\it all} length scales, and span from ballistic to diffusive transport regimes. The transmission coefficient controls the length dependence of the heat current. In the ballistic limit $L\ll\lambda$, $\mathcal{T}\rightarrow 1$ and $I_Q$ is independent of length. Note that in this limit $I_Q$ is equal to the known expression in the Casimir limit (see Appendix \ref{sec:casimir}). In the diffusive limit $L\gg\lambda$, $\mathcal{T}\approx \lambda/L$ and $I_Q\propto 1/L$, as expected from classical scaling. Eq. (\ref{fnet2}) is identical to $I_Q$ obtained with the Landauer approach \cite{Kim2009,Jeong2011}. One advantage of the McKelvey-Shockley flux method, versus Landauer, is that it provides spatial information on the heat transport properties.

Eq. (\ref{def:heat_den}) shows that the heat density is the sum $I_Q^+ + I_Q^-$ divided by the average forward projected velocity $v_x^+$. Often it is desirable to replace heat density by temperature. It is important to note that in small structures (compared to $\lambda$) the phonon distribution may be highly non-equilibrium and the definition of temperature, an equilibrium quantity, is ambigious (no such problem arises with $I_Q$ and $Q$). Assuming a small applied $\Delta T = T_L - T_R$ ensures the thermal conductor remains near equilibrium, where temperature is well defined. In this case, we can rewrite heat density in terms of temperature using $\delta Q = C_V\,\delta T$, where $C_V$ is the volumetric heat capacity. And, we can expand the heat currents $I_Q^{\pm}(x)=I_{Q,\rm eq}^{\pm} + \delta I_Q^{\pm}(x)$ due to a small difference in contact temperatures $\Delta T$, where $I_{Q,\rm eq}^+=I_{Q,\rm eq}^-$ is simply the equilibrium heat current arising from a reference background temperature, chosen as $T_R$ in this case. Rewriting $Q$ in terms of $T$, we find:
\begin{align}
\delta T^{\pm}(x) &= \frac{2\delta I_Q^{\pm}(x)}{C_V\,v_x^+}, \label{def:temp_pm} \\
T(x) &= \left[\frac{\delta T^+(x) + \delta T^-(x)}{2}\right] + T_R. \label{def:temp}
\end{align}
The temperature profile inside the thermal conductor is proportional to the sum of $\delta I_Q^+$ and $\delta I_Q^-$ or equivalently the average of $\delta T^+(x)$ and $\delta T^-(x)$, and is presented in Fig. \ref{fig3}(c). Note that $\delta Q^{\pm}=(C_V/2) \,\delta T^{\pm}$, since $C_V$ takes into account both the forward and backward phonon states. Using this with the property $\delta Q = \delta Q^+ + \delta Q^-$, we obtain the above expression for $T(x)$, which has the appearance of an average over $\delta T^{\pm}$. We see that $T(x)$ varies linearly inside the thermal conductor, and that there are discrete temperature drops at the boundaries. The separation of temperature for phonons traveling in the forward and backward directions is completely analogous to the way electrochemical potentials are treated with electron transport in nanostructures \cite{Mclennan1991}.

Allowing the forward- and backward-moving phonon populations, and associated temperatures ($T^+$ and $T^-$), to be different is key to capturing the non-equilibrium nature of ballistic transport. In the diffusive limit the difference between $T^+$ and $T^-$, as well as $I_Q^+$ and $I_Q^-$, becomes vanishingly small, and corresponds to the case of near local thermal equilibrium (i.e. both halves of the phonon distribution are nearly identical).

We can use Eqns.(\ref{def:temp_pm}-\ref{def:temp}) with Eqns.(\ref{fplus}-\ref{fminus}) to determine the temperatures at the boundaries:
\begin{align}
T(0^+) &= (2-\mathcal{T})\,\frac{T_L}{2} + \mathcal{T}\,\frac{T_R}{2}, \label{temp_left} \\
T(L^-) &= \mathcal{T}\,\frac{T_L}{2} + (2-\mathcal{T})\,\frac{T_R}{2}, \label{temp_right}
\end{align} 
where the relation $T_L = T_R +2 \delta I_{Q,0}^+/ C_Vv_x^+$ was used (the factor of two appears since the forward and backward heat currents are equal in the contacts). The boundary temperatures $T(0^+)$/$T(L^-)$ are weighted averages of the contact temperatures, that depend on $\mathcal{T}$. When $\mathcal{T}\rightarrow 0$ (diffusive limit), the ``interior" boundary temperatures tend to the contact temperatures (classical result), and when $\mathcal{T}\rightarrow 1$ (ballistic limit) the boundary temperatures tend to the average of the contact temperatures (constant $T(x)$ inside the material) \cite{Chen2000}. From Eqns.(\ref{temp_left}-\ref{temp_right}), we can extract the value of the temperature jumps at the contacts ($\delta T_c$). $\delta T_c$ is found to be identical at both left and right contacts:
\begin{align}
\delta T_c &= \frac{\mathcal{T}}{2}\left(T_L-T_R\right), \label{temp_jump} \\
           &= \frac{I_QR^{\rm ball}}{2}, \label{temp_jump2}
\end{align} 
where $R^{\rm ball}=1/K^{\rm ball}$ is the ballistic thermal resistance with $K^{\rm ball} = C_Vv_x^+/2$. We find the temperature jumps are simply proportional to the transmission, which depends on $L$ and $\lambda$, and can be interpreted as an intrinsic contact resistance. The temperature jumps do not occur because the contact interfaces scatter phonons, since we assume reflectionless contacts. Rather, it is because we specify the injected heat currents $I_{Q,0}^+$/$I_{Q,L}^-$ at one end of the thermal conductor while the opposite ends are ``floating" boundaries that depend on $\lambda$ and $L$ (as shown in Fig. \ref{fig3}(b)). Once the forward and backward heat currents are added to obtain temperature, discrete drops are observed at the boundaries.

In this work we have assumed perfectly thermalizing contacts that are in equilibrium, which allows us to extract a simple analytical expression for the temperature jumps. In general one must specify the injected heat currents at the boundaries, which in principle can originate from an adjacent material that need not be in equilibrium.

\begin{figure}	
\includegraphics[width=6cm]{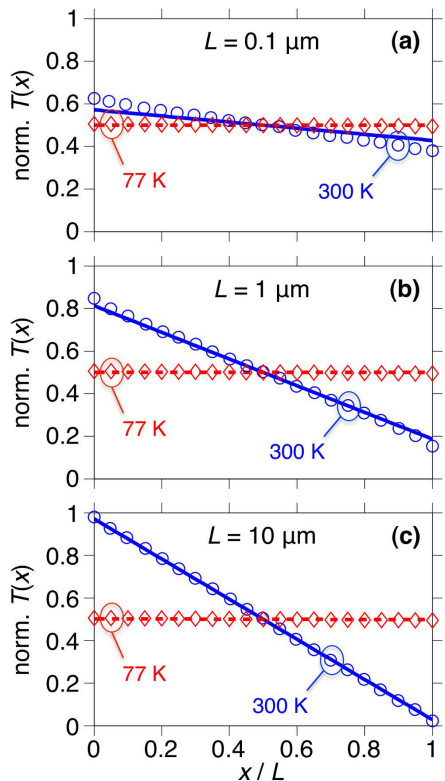}
\caption{(Color online) Normalized temperature profile $(T(x)-T_R)/(T_L-T_R)$ versus normalized position $x/L$ of diamond films of length $L$ = $0.1\,{\rm \mu m}$ (a), $1\,{\rm \mu m}$ (b), $10\,{\rm \mu m}$ (c). Lines correspond to our analytical solution, and symbols are obtained from the phonon BTE (taken from \cite{Majumdar1993}). The values $\lambda=2840\,{\rm \mu m}$ ($77\,{\rm K}$) and $\lambda=596\,{\rm nm}$ ($300\,{\rm K}$) reported in \cite{Majumdar1993} were adopted.} 
\label{fig4}
\end{figure}

In Fig. \ref{fig4} we compare our simple model (lines) to the numerical solutions of the BTE (symbols), in the case of diamond films of length $L$ = $0.1\,{\rm \mu m}$, $1\,{\rm \mu m}$, $10\,{\rm \mu m}$. Using the effective $\lambda$ values reported in \cite{Majumdar1993} for $T=77,300 \,{\rm K}$, we find our simple analytical model adequately reproduces the results of the BTE, including the linear temperature profiles and the temperature jumps at the contacts. We note that phonons in a material typically have a broad distribution of $\lambda$, and in general using a constant $\lambda$ may lead to significant errors (in this case $\delta T_c$ should be obtained by evaluating $\lambda$ and $\mathcal{T}$ at each energy and integrating over energy following Eqns.(\ref{def:heat_curr}-\ref{def:heat_den}), as discussed in Section \ref{sec:discussion}).

To summarize, writing the phonon BTE in the McKelvey-Shockley form (Eqns.(\ref{mk_flux1}) and (\ref{mk_flux2})) leads to very accurate solutions for the class of problems considered in this section. In the next section, we will show that exactly the same solutions can be obtained by solving the traditional diffusion equation -- if the appropriate boundary conditions are used.

\section{Fourier's Law and Heat Equation} 
\label{sec:fourier}
In this section, we demonstrate how the McKelvey-Shockley flux equations can be rewritten into familiar expressions for heat transport, that yield exactly the same results. By adding Eq. (\ref{mk_flux1}) and Eq. (\ref{mk_flux2}) (replacing $F^{\pm}$ by $I_Q^{\pm}$) and using the relation $Q=(I_Q^+ + I_Q^-)/v_x^+$, we obtain
\begin{align}
I_Q &= -D_{\rm ph} \, \frac{{\rm d} Q(x)}{{\rm d} x} \label{fourier1} \\
      &= -\kappa \, \frac{{\rm d} T(x)}{{\rm d} x}, \label{fourier2}
\end{align}
where 
\begin{align}
D_{\rm ph} &= \lambda \, v_x^+/2, \label{diff_coef} \\
\kappa &= C_V D_{\rm ph}, \label{kappa_bulk}
\end{align}
$D_{\rm ph}$ is the phonon diffusion coefficient and $\kappa$ is the bulk thermal conductivity (see Appendix \ref{sec:trad_kappa}). Eq. (\ref{fourier2}) is Fourier's law, and comes out directly from the McKelvey-Shockley flux method, {\it without making any assumption on $L$ relative to $\lambda$}. This indicates that the ballistic transport physics contained in the McKelvey-Shockley flux method are also included in Fourier's law. By combining Fourier's law with the energy balance equation (Eq. (\ref{cont_eq})), we find the steady-state heat equation:
\begin{align}
\frac{{\rm d}^2Q(x)}{{\rm d}x^2} &= \frac{{\rm d}^2T(x)}{{\rm d}x^2} = 0, \label{heat_eq}
\end{align}
where we assumed the material parameters are position-independent. Fourier's law has been derived from the BTE \cite{Majumdar1993}, by assuming that the phonons at each $x$ were locally at thermodynamic equilibrium. We find no such assumption is necessary. The key to capturing ballistic transport effects with Fourier's law is to use the correct physical boundary conditions.

Traditionally, the contact temperatures $T_L$/$T_R$ are used as the boundary conditions, however the McKelvey-Shockley flux method shows it is the injected heat currents that are the physical boundary conditions (Eqns.(\ref{flux_bound1}-\ref{flux_bound2})). Stated differently, the BTE is a first order equation in space, and requires one boundary condition for the phonon distribution. This is typically accomplished by defining a boundary condition on half of the distribution (the incoming flux) at each of the two contacts. Specifying the contact temperatures as the boundary conditions at both ends overspecifies the problem. The temperature at each of the two ends is determined by the temperatures of the injected fluxes and by the scattering within the film; it is an outcome of the calculation. Although, in the diffusive limit, the boundary temperatures tend asymptotically to the contact temperatures. In general, using the definitions of heat current and heat density, the boundary conditions for temperature are determined to be (details in Appendix \ref{sec:bc_temp}):
\begin{align}
2\,\delta I_{Q,0}^+ &= -\kappa \left.\frac{{\rm d}(\delta T)}{{\rm d}x}\right|_{0^+} + C_Vv_x^+\delta T(0^+), \label{bc_temp1} \\
2\,\delta I_{Q,L}^-  &= +\kappa \left.\frac{{\rm d}(\delta T)}{{\rm d}x}\right|_{L^-} + C_Vv_x^+\delta T(L^-), \label{bc_temp2}
\end{align}
where $T(x)=\delta T(x) + T_R$, $\delta T(x) = [\delta T^+(x) + \delta T^-(x)]/2$ and $T_R$ was chosen as our reference background temperature. The above boundary conditions are mixed, meaning they depend on both temperature and its derivative, and are applicable even when the contacts are not in equilibrium.

Solving the heat equation (Eq. (\ref{heat_eq})) is straightforward, and simply gives a linear temperature profile
\begin{align}
T(x) &= T(0^+)\,\left[1-\frac{x}{L}\right] + T(L^-)\,\left[\frac{x}{L}\right]. \label{temp_profile}
\end{align} 
Inserting this solution into Eqns.(\ref{bc_temp1}-\ref{bc_temp2}), we can determine $T(0^+)$ and $T(L^-)$, which in this case (equilibrium contacts) are equal to Eqns.(\ref{temp_left}-\ref{temp_right}). With equilibrium contacts, applying the above boundary conditions is equivalent to replacing the contact temperatures $T_L$ and $T_R$ by the ``interior" boundary temperatures $T(0^+)$ and $T(L^-)$. We note that using Eqns.(\ref{temp_left}-\ref{temp_right}) with Eq. (\ref{temp_jump2}), $T(0^+)$ and $T(L^-)$ can also be rewritten as 
 \begin{align}
T(0^+) &= T_L - \frac{R^{\rm ball}I_Q}{2}, \label{temp_left2} \\
T(L^-) &= T_R + \frac{R^{\rm ball}I_Q}{2}. \label{temp_right2}
\end{align}

With a given $T(x)$, we can calculate the heat current using Fourier's law (Eq. (\ref{fourier2})):
\begin{align}
I_Q &= \kappa \, \left[\frac{T(0^+) - T(L^-)}{L} \right], \label{heat_curr1}\\
      &= \kappa \, \left[\frac{T_L - T_R}{L+\lambda} \right]. \label{heat_curr2}
\end{align}
Note the correct usage of Fourier's law implies  evaluating the gradient of $T$ using the ``interior" boundary temperatures and not the contact temperatures. This expression for heat current is applicable on {\it all} length scales, and is equal to Eq. (\ref{fnet3}). In the diffusive limit ($L\gg\lambda$), we have $I_Q=\kappa \, (T_L - T_R)/L$, the classical result; in the ballistic limit ($L\ll\lambda$), we have $I_Q=\kappa \, (T_L - T_R)/\lambda$ and the heat current is independent of $L$. Eq. (\ref{heat_curr2}) shows that $|{\rm d}T/{\rm d}x|$ is reduced, due to the temperature jumps at the boundaries, and is equivalent to replacing $L$ by $L+\lambda$. As transport becomes more ballistic, the temperature jumps increase and the absolute temperature gradient inside the thermal conductor decreases. If one assumed ${\rm d}T/{\rm d}x=(T_L-T_R)/L$ for all $L$, then a reduction in the expected $I_Q$ could be interpreted as a reduction in the thermal conductivity. This introduces the concept of apparent thermal conductivity, which is mathematically defined as $\kappa_{\rm app}=\kappa \cdot L/(L+\lambda)$ \cite{Jeong2011}. Our analysis shows that, physically, ballistic transport reduces ${\rm d}T/{\rm d}x$ and does not change the bulk thermal conductivity of the material at any length scale -- as long as $\lambda$ is independent of energy.

\section{Role of full phonon dispersion and mean-free-path distribution}
\label{sec:discussion}
\begin{figure}	
\includegraphics[width=6cm]{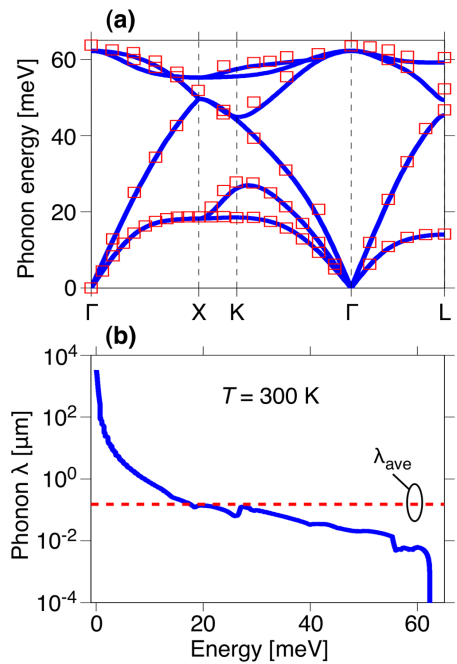}
\caption{(Color online) (a) Phonon dispersion of bulk silicon computed from first principles (lines: theory; symbols: experimental data \cite{Nilsson1972}). (b) Mean-free-path for backscattering $\lambda$ versus energy at room temperature. $\lambda$ includes boundary, defect and Umklapp scattering, and was calibrated to experimental data. Dashed horizontal line corresponds to the average bulk $\lambda_{\rm ave}=151\,{\rm nm}$.} 
\label{fig5}
\end{figure}

In Section \ref{sec:example}, we validated our approach by comparing our solutions to the numerical results of the BTE where a constant $\lambda$ was used. In general, the MFP of phonons in a given material can span orders of magnitude \cite{Henry2008}, and may not be well represented by a single MFP. Next, we investigate the impact of an energy-dependent $\lambda(\epsilon)$, and demonstrate how to perform detailed modeling using the ``simple" approach presented in this paper. Specifically, we calculate the temperature profile inside a silicon film, using a first principles-calculated phonon dispersion and a semi-empirical scattering model calibrated to experimental data. From the full phonon dispersion, shown in Fig. \ref{fig5}(a) (details in \cite{SiPhonon}), we obtain $M(\epsilon)$ and $v_x^+(\epsilon)$. Including Umklapp, defect and boundary scatterings, we extract $\lambda(\epsilon)$, presented in Fig. \ref{fig5}(b) (details in \cite{SiMFP}). With this model, we can reproduce the lattice thermal conductivity of bulk silicon within 15\% error from $5\,{\rm K}$ to $300\,{\rm K}$.

Fig. \ref{fig1} shows the normalized temperature profile versus normalized position of Si films for $L=3,\,30,\,300\,{\rm nm}$. Our solutions (lines) are compared to results of the phonon BTE (symbols) \cite{Escobar2008}. Given that small differences would be expected since we do not use the exact same dispersion and MFP distribution as in Ref. \cite{Escobar2008}, the agreement is very good. If the average bulk $\lambda$ is used ($\lambda_{\rm ave}=151\,{\rm nm}$), instead of the energy-dependent $\lambda(\epsilon)$, significant differences in the temperature profiles are observed. For example, with $\lambda(\epsilon)$ the temperature jumps are $0.43$ ($L=3\,{\rm nm}$), $0.27$ ($L=30\,{\rm nm}$) and $0.09$ ($L=300\,{\rm nm}$), however with $\lambda_{\rm ave}$ we find $0.49$, $0.42$ and $0.17$, respectively. Thus, using $\lambda_{\rm ave}$ overestimates the phonon ballisticity, because it fails to capture the fact that a large fraction of the phonons have a MFP less than the average (see Fig. \ref{fig5}(b)).

How does the temperature profile change when using an energy-dependent $\lambda(\epsilon)$? According to Eq. (\ref{def:heat_den}) and the definition of heat capacity $\delta Q(x,\epsilon) = C_V(\epsilon)\, \delta T(x,\epsilon)$, the energy-averaged temperature at any point $x$ is given by:
\begin{align}
\delta T(x) &= \frac{\int_0^{\infty}C_V(\epsilon)\,\delta T(x,\epsilon)\,{\rm d}\epsilon}{\int_0^{\infty}C_V(\epsilon)\,{\rm d}\epsilon},\label{temp_total}
\end{align}
where
\begin{align} 
C_V(\epsilon) &= \epsilon \left[\frac{2M(\epsilon)}{h\,v_x^+(\epsilon)}\right] \frac{\partial f_{\rm BE}(\epsilon)}{\partial T}. \label{cv_energy}
\end{align}
Note that $C_V = \int_0^{\infty} C_V(\epsilon)\,{\rm d}\epsilon$, and that the term in brackets is simply the phonon density of states \cite{Jeong2011}. From the temperature profile given by Eq. (\ref{temp_profile}), we only need to compute the energy-averaged temperatures $T(0^+) = \delta T(0^+)+T_R$ and $T(L^-) = \delta T(L^-)+T_R$ with Eq. (\ref{temp_total}). A straight line connecting both points corresponds to the correct energy-averaged temperature profile. $\delta T(0^+,\epsilon)$ and $\delta T(L^-,\epsilon)$ are both related to the temperature jump $\delta T_c(\epsilon)$ at the contacts (Eq. (\ref{temp_jump})) and thus $\lambda(\epsilon)$. Hence, if $M(\epsilon)$, $v_x^+(\epsilon)$ and $\lambda(\epsilon)$ are known (using full phonon dispersions and energy-dependent scattering rates), performing detailed modeling within the simple approach described in this paper boils down to evaluating one integral in energy to extract the energy-averaged $\delta T$ or $\mathcal{T}$. From this the temperature profile corresponds to a straight line joining $T(0^+)$ and $T(L^-)$. Note that if $\lambda$ is a constant, then $\delta T$ does not depend on energy, evaluating Eq. (\ref{temp_total}) is trivial, and $M(\epsilon)$ and $v_x^+(\epsilon)$ do not need to be specified.

\section{Summary}
\label{sec:summary}
In summary, using the McKelvey-Shockley flux method, we have shown that a simple phonon Boltzmann equation can be written. The solutions to this equation agree well with the results of the full phonon BTE, including temperature jumps ($\delta T_c$) at the boundaries with ideal contacts. Examples with a simple phonon dispersion and energy-independent phonon mean-free-path and with a full band phonon dispersion and energy-dependent mean-free-path distribution were both considered. For the simple case, analytical solutions for $T(x)$ and $I_Q$, that describe phonon transport from the ballistic to diffusive limits were derived. For the complicated case, the results of full numerical solutions of the phonon Boltzmann equation can be reproduced with a fraction of the computational burden.

In addition to faster solutions, the method introduced here also provides new insights into quasi-ballistic phonon transport. For example, the nature of $\delta T_c$ was discussed, and we showed that it is simply related to the phonon transmission across the film. In addition, we showed that $\delta T_c$ can be described in terms of an ideal contact resistance equal to one-half of the ballistic thermal resistance at each contact, which is analogous to the so-called quantum contact resistance for electron transport \cite{Datta1997}. We also showed that for a constant mean-free-path, the thermal conductivity of a small structure ($\lambda > L$) is the same as the bulk thermal conductivity. The reduction in heat flux occurs because the temperature difference across the film is less than the difference in temperatures of the two contacts and not, as commonly modeled, because of reduced thermal conductivity.

Finally, we showed that our simple phonon Boltzmann equation can be rewritten exactly as Fourier's law and the heat equation. When solved with correct boundary conditions, we showed that Fourier's law and the heat equation capture ballistic effects and are thus applicable on all length scales.

This work addressed one-dimensional transport where the approach and effects can be most clearly discussed. If extensions to higher dimensions are similarly accurate, this approach may prove useful in extending finite-element heat transfer tools to capture ballistic effects and analyzing realistic structures and experiments probing short length and time scales, such as time/frequency-domain thermoreflectance. For the latter, a time-dependent McKelvey-Shockley flux method of the type previously used to described electron transport \cite{Alam1995} will be needed.

\acknowledgements
This work was supported in part by DARPA MESO (Grant N66001-11-1-4107) and through the NCN-NEEDS program, which is funded by the National Science Foundation, contract 1227020-EEC, and the Semiconductor Research Corporation. J.M. acknowledges financial support from NSERC of Canada.

\appendix
\section{Heat current in the Casimir limit}
\label{sec:casimir}
In this appendix we show how our derived expression for heat current (Eq. (\ref{fnet2})), in the ballistic phonon limit, reduces to the known result in the Casimir limit. Starting from Eq. (\ref{fnet2}), we assume purely ballistic transport, meaning $\lambda \gg L$ and $\mathcal{T}\rightarrow 1$, which gives
\begin{align}
I_Q^{\rm tot} &= \int_0^{\infty}\epsilon\,\frac{M(\epsilon)}{h} \left[f_{\rm BE}(\epsilon,T_L) - f_{\rm BE}(\epsilon,T_R)\right]  \,{\rm d}\epsilon, \label{app:IQ_ball}
\end{align}
where we have integrated over energy to obtain the total heat current (using Eq. (\ref{def:heat_curr})). With a linear phonon dispersion, as is commonly assumed in the Casimir limit, the distribution of modes is \cite{Jeong2011}:
\begin{align}
M(\epsilon) &= \frac{3 \pi \epsilon^2}{h^2 v_g^2}, \label{app:dom_ph}
\end{align}
where $v_g$ is the group velocity and the factor of three is for the three acoustic branches. Inserting Eq. (\ref{app:dom_ph}) and the expression for $f_{\rm BE}$ into Eq. (\ref{app:IQ_ball}), we find
\begin{align}
I_Q^{\rm tot} &= \frac{3 \pi}{h^3 v_g^2} \int_0^{\infty}\!\! \left[\frac{\epsilon^3}{{\rm e}^{\epsilon/k_B T_L}-1} - \frac{\epsilon^3}{{\rm e}^{\epsilon/k_B T_R}-1}\right] \,{\rm d}\epsilon, \label{app:IQ_ball2}
\end{align}
where $k_B$ is Boltzmann's constant. Defining a new variable $y=\epsilon/k_B T_{L,R}$, Eq. (\ref{app:IQ_ball2}) becomes:
\begin{align}
I_Q^{\rm tot} &= \frac{3 \pi k_B^4}{h^3 v_g^2} \left( T_L^4-T_R^4 \right) \int_0^{\infty}\!\! \frac{y^3}{{\rm e}^y-1} \,{\rm d}y. \label{app:IQ_ball3}
\end{align}
One can show that the integral is equal to $\pi^4/15$, which gives the known result for heat current in the Casimir limit \cite{Majumdar1993}:
\begin{align}
I_Q^{\rm tot} &= \sigma \left( T_L^4-T_R^4 \right), \label{app:IQ_casimir} \\
\sigma &= \frac{\pi^5 k_B^4}{5 h^3 v_g^2}, \label{app:stefan_boltz}
\end{align}
where $\sigma$ is the Stefan-Boltzmann constant for phonons. Note the above expression is valid at temperatures much less than the Debye temperature of the material.

\section{Traditional expression for thermal conductivity}
\label{sec:trad_kappa}
The expression for bulk thermal conductivity derived from the approach presented in this work is (see Eq. (\ref{kappa_bulk})): 
\begin{align}
\kappa &= C_V \lambda\, v_x^+ /2, \label{app:kappa_bulk}
\end{align}
where $C_V$ is the heat capacity, $\lambda$ the mean-free-path {\it for backscattering} and $v_x^+$ is the average $x$-projected velocity of the forward moving carriers. The commonly encountered relation for thermal conductivity is 
\begin{align}
\kappa &= C_V l\, v_g /3, \label{app:trad_kappa_bulk}
\end{align}
where $l$ is the mean-free-path and $v_g$ the group velocity. We can show that both Eq. (\ref{app:kappa_bulk}) and Eq. (\ref{app:trad_kappa_bulk}) are identical, in the case of an isotropic phonon dispersion (note our approach applies in the case of any full phonon dispersion). In \cite{Jeong2010}, it is shown that $v_x^+ =v_g/2$ and $\lambda = (4/3)\,l$. Inserting these relations into Eq. (\ref{app:kappa_bulk}), one directly finds Eq. (\ref{app:trad_kappa_bulk}).

\section{Boundary conditions for temperature}
\label{sec:bc_temp}
In this note we demonstrate how to derive Eqns. (\ref{bc_temp1}-\ref{bc_temp2}). Our starting point is Fourier's law, which we showed is applicable on all length scales (Eq. (\ref{fourier2})):
\begin{align}
I_Q^+(x)-I_Q^-(x) &= -\kappa \frac{{\rm d}T(x)}{{\rm d}x}, \label{app:net_heat_curr}
\end{align}
where $I_Q^+(x)-I_Q^-(x)$ is simply the net heat current $I_Q$. Using the expression for heat density (Eq. (\ref{def:heat_den})), we have
\begin{align}
I_Q^+(x)+I_Q^-(x) &= v_x^+ Q(x). \label{app:heat_den}
\end{align}
By adding Eqns. (\ref{app:net_heat_curr}) and (\ref{app:heat_den}) and evaluating $x$ at the left boundary ($x=0^+$), we obtain
\begin{align}
2 I_{Q,0}^+ &= -\kappa \left.\frac{{\rm d}T}{{\rm d}x}\right|_{0^+} + v_x^+ Q(0^+). \label{app:left_bc_temp}
\end{align}
By subtracting Eqns. (\ref{app:net_heat_curr}) from (\ref{app:heat_den}) and evaluating $x$ at the right boundary ($x=L^-$), we obtain
\begin{align}
2 I_{Q,L}^- &= \kappa \left.\frac{{\rm d}T}{{\rm d}x}\right|_{L^-} + v_x^+ Q(L^-). \label{app:right_bc_temp}
\end{align}
Since we assume the applied temperature difference across the contacts ($\Delta T = T_L-T_R$) is small, the heat currents can be expanded as $I_Q^{\pm}(x) = \delta I_Q^{\pm}(x) + I_{Q,{\rm eq}}^{\pm}$, where $I_{Q,{\rm eq}}^+=I_{Q,{\rm eq}}^-$ is the equilibrium heat current associated with a reference background temperature (chosen as $T_R$ in this case) and $\delta I_Q^{\pm}(x)$ is a correction due to $\Delta T$. Rewriting Eqns. (\ref{app:left_bc_temp}) and (\ref{app:right_bc_temp}) in terms of the injected heat fluxes due to $\Delta T$ and using the definition for heat capacity $C_V$ ($\delta Q=C_V \delta T$), we find: 
\begin{align}
2\, \delta I_{Q,0}^+ &= -\kappa \left.\frac{{\rm d}(\delta T)}{{\rm d}x}\right|_{0^+} + C_V v_x^+ \delta T(0^+), \label{app:left_bc_temp2} \\
2\, \delta I_{Q,L}^- &= \kappa \left.\frac{{\rm d}(\delta T)}{{\rm d}x}\right|_{L^-} + C_V v_x^+ \delta T(L^-), \label{app:right_bc_temp2}
\end{align}
where $T(x)=\delta T(x) + T_R$. The above equations relate the injected heat currents at the boundaries to the temperature (and its gradient) at the boundaries, and represent the correct physical boundary conditions for the heat equation and Fourier's law.

Note that Eqns. (\ref{app:left_bc_temp2}-\ref{app:right_bc_temp2}) are applicable even when the contacts are not in equilibrium. If the contacts are in equilibrium, then $\delta I_{Q,0}^+$ and $\delta I_{Q,L}^-$ can be obtained by expanding Eqns. (\ref{fleft}-\ref{fright}):
\begin{align}
I_{Q,0}^+ &= \epsilon\,\frac{M}{h}\,\left[f_{\rm BE}(T_R)+\frac{\partial f_{\rm BE}}{\partial T}\Delta T\right] \label{app:left_inj_flux} \\
                &= I_{Q,{\rm eq}}^+ + \delta I_{Q,0}^+, \nonumber \\
I_{Q,L}^-  &= \epsilon\,\frac{M}{h}\,f_{\rm BE}(T_R) \label{app:right_inj_flux} \\
                &= I_{Q,{\rm eq}}^-. \nonumber
\end{align}
By choosing $T_R$ as the reference temperature, we have $\delta I_{Q,L}^-=0$.

%\bibliographystyle{}
%\nocite{*}


\begin{thebibliography}{00}
\bibitem{SiPhonon} The Si phonon dispersion was computed within the harmonic approximation, with the force constants obtained from DFPT as implemented in VASP \cite{VASP}. A $3$$\times$$3$$\times$$3$ cubic supercell containing Si 216 atoms was used (lattice constant of $5.47\,{\rm \AA}$). A plane-wave energy cutoff of $350\,{\rm eV}$, a $3$$\times$$3$$\times$$3$ $k$-mesh (Monkhorst-Pack scheme) were tested to provide good accuracy. The PAW method and GGA-PBE were adopted for the core and exchange-correlation potentials, respectively. The phonon dispersion was obtained, from the force constants, using the Phonopy simulation package \cite{Phonopy}.
\end{thebibliography}
\end{document}